\documentclass{moriond}
\usepackage{amssymb}
\usepackage{amsmath}
\usepackage{amscd}
\usepackage{latexsym}
\usepackage{slashed}
\usepackage{color}
\usepackage{graphicx}
\usepackage[normalem]{ulem}
\usepackage{color}
\definecolor{orange}{cmyk}{0,0.5,1,0}
\usepackage{verbatim}
\usepackage{rotating}
\usepackage{braket}
\usepackage{diagbox}

\def\lsim{\raise0.3ex\hbox{$\;<$\kern-0.75em\raise-1.1ex\hbox{$\sim\;$}}}
\def\gsim{\raise0.3ex\hbox{$\;>$\kern-0.75em\raise-1.1ex\hbox{$\sim\;$}}}
\usepackage{graphicx,multirow,subfigure}
\usepackage{caption}

\def\be{\begin{equation}}
\def\ee{\end{equation}}
\def\bea{\begin{eqnarray}}
\def\eea{\end{eqnarray}}

\def\be{\begin{equation}}
\def\ee{\end{equation}}
\def\bea{\begin{eqnarray}}
\def\eea{\end{eqnarray}}

\newcommand{\Photo}{}

\begin{document}
\vspace*{-1cm}
\title{${}^{8}$Be Decay Anomaly and Light $Z'$}

\author{Luigi Delle Rose,}
\address{School of Physics and Astronomy, University of Southampton, Highfield, Southampton SO17 1BJ, United Kingdom }
\address{INFN, Sezione di Firenze, and Dipartimento di Fisica ed Astronomia,	Universit\`a di  Firenze, Via G. Sansone 1, 50019 Sesto Fiorentino, Italy}
\author{Shaaban Khalil,}
\address{Center for Fundamental Physics, Zewail City of Science and Technology, Sheikh Zayed,12588 Giza, Egypt}
\author{\textbf{Simon J.D. King}\footnote{Speaker},}
\address{School of Physics and Astronomy, University of Southampton, Highfield, Southampton SO17 1BJ, United Kingdom }
\address{INFN, Sezione di Padova, and Dipartimento di Fisica ed Astronomia G. Galilei, Universit\`a di Padova, 
Via Marzolo 8, 35131 Padova, Italy}
\author{Stefano Moretti}
\address{School of Physics and Astronomy, University of Southampton, Highfield, Southampton SO17 1BJ, United Kingdom }
\address{Particle Physics Department, Rutherford Appleton Laboratory, Chilton, Didcot, Oxon OX11 0QX, United Kingdom}
\author{Ahmed M. Thabt}
\address{Center for Fundamental Physics, Zewail City of Science and Technology, 6 October City, Giza 12588, Egypt}

\maketitle
\abstracts{In this proceedings, we discuss a light (17 MeV) $Z'$ solution to the anomaly observed in the decay of Beryllium-8 by the Atomki collaboration. We detail an anomaly free model with minimal particle content which can satisfy all other experimental constraints with gauge couplings $\mathcal{O}(10^{-4})$. }

\section{Introduction}
In the quest to find New Physics (NP), there is another possible route besides the traditional high energy and precision frontiers, which is the low energy regime. There has been a consistent anomaly in the measured angular distribution of $e^+ e^-$ pairs in the 18.15 MeV decay of the excited state Be$^*$, by the Atomki collaboration over the last few years \cite{Krasznahorkay:2017gwn,Krasznahorkay:2017bwh,Krasznahorkay:2017qfd,Krasznahorkay:2018snd}. In the Standard Model (SM), at such low transition energies, this process is mediated by a photon with a Branching Ratio (BR) $\approx 1$, and, due to this massless mediator, one expects few events at large angles between the electron and positron. However, what is seen by the Atomki collaboration is an excess of events at large angles $\sim 140^{\circ} $. The simplest explanation is to introduce a new boson with mass just under the transition energy, such that it is produced close to on-shell, which then decays to the $e^{+}e^{-}$ pair which is detected by the apparatus. This boson can be either of vector (spin 1) or of pseudoscalar (spin 0) nature \cite{DelleRose:2018pgm}. 
The Atomki collaboration made an initial determination that the best mass and BR of such a boson should be

\begin{equation}
M_{Z'} = 16.7 \pm 0.35 \textrm{(stat)} \pm 0.5 \textrm{(sys)} \textrm{ MeV},
\end{equation}

\begin{align}
\label{eq:Br}
R  &\equiv {\frac{{\rm BR}(^8{{\rm Be}^*} \to Z' + {^8{\rm Be}})}{{\rm BR}(^8{{\rm Be}^*} \to \gamma + {^8{\rm Be}})} \times {\rm BR}(Z'\to e^+ e^-)} =5.8 \times 10 ^{-6} .
\end{align}
Subsequently, additional masses and BR combinations were suggested in a private communication to Feng {\sl et al.} \cite{Feng:2016ysn}, which we list in Tab. \ref{tab:Br}, and eventually consider all possibilities in our results.
\begin{table}[!t]
\centering
\begin{tabular}{c | c}
	$M_{Z'} ~(\textrm{MeV}) $ & $R$ \\ \hline \vspace{-1em}
	&\\ 
	16.7 & $5.8 \times 10^{-6}$ \\
	17.3 & $2.3 \times 10^{-6}$ \\
	17.6 & $5.0 \times 10^{-7}$
\end{tabular}
\caption{Solutions to the Atomki anomaly, with best fit mass value (16.7 MeV), 
and subsequent alternative masses (17.3 MeV and 17.6 MeV), 
along with the corresponding ratio of BRs, $R$, as defined in Eq. (\ref{eq:Br}).}
\label{tab:Br}
\end{table}

In this work, we will focus on a vector boson explanation to match these conditions for the anomaly explanation and construct a minimal scenario which may evade all other experimental constraints \cite{DelleRose:2018eic}. We first extend the SM by a new $U(1)'$ group. After rotating away the new mixed term in the kinetic Lagrangian, one finds a covariant derivative
\begin{equation}
{\cal D}_\mu = \partial_\mu + .... + i g_1 Y B_\mu + i (\tilde{g} Y + g' z) B'_\mu,
\end{equation}
where $z$ defines the charge of the field content under the new $U(1)'$ group, with associated gauge boson $B'$, and $g',\tilde{g}$ are the new gauge coupling and gauge-kinetic mixing, respectively. The gauge boson interacts with the fermions through the gauge current 
\begin{equation}
J^\mu_{Z'} = \sum_f \bar \psi_f \gamma^\mu \left( C_{f, L} P_L + C_{f, R} P_R \right) \psi_f ,
\end{equation}
and, assuming the limit of small gauge coupling and mixing, we find the vector and axial couplings
\begin{eqnarray}
C_{f, V} &=& \frac{C_{f,R} + C_{f,L}}{2}\nonumber\\
	&\simeq &\tilde g  c_W^2 \, Q_f + g' \left[ z_H (T^3_f - 2 s_W^2 Q_f)  + z_{f,V} \right], \\
C_{f, A} &=& \frac{C_{f,R} - C_{f,L}}{2} \simeq  g' \left[   -  z_H \, T^3_f  +   z_{f,A} \right],
\label{eq:Axial}
\end{eqnarray}
where we introduce the notation $z_{f,V/A}=(z_{f_R} \pm z_{f_L})/2$, $s_W = \sin \theta _W$ and  $c_W = \cos \theta _W$.

Assuming a family non-universal scenario, one has 15 free parameters, the 12 SM fermion charges under the new group, the SM Higgs charge, new gauge coupling, $g'$, and gauge-kinetic mixing parameter, $\tilde{g}$. In the next sections we detail motivations to fix the field charges from some reasonable requirements of the theory.

\section{Charge Assignment}
We first require that the Atomki condition itself be satisfied. Though we do not detail the nuclear physics here, this requires that up and down quarks have non-zero axial couplings to the gauge boson, $C_{u/d,A} \neq 0$.

We also demand that the theory be anomaly free without requiring extra vector-like states, and the six possible triangle diagrams be satisfied by only SM + Right Handed (RH) neutrinos \footnote{We see in our particular set-up that at least two RH neutrinos are required to be charged under the new $U(1)'$.}.

Next, we subject our scenario to constraints from experimental data. Introducing a new $Z'$, even with small gauge couplings, is very sensitive to certain experiments. Firstly, neutrino experiments are very constraining, such as meson decays like $K^{\pm}\rightarrow \pi ^{\pm} \nu \nu$ and electron-neutrino scattering from the TEXONO experiment. Since the $Z'$ must couple to electrons, any neutrino coupling will show clear deviations from such precise experiments. To this end, we require no vector or axial couplings to the neutrinos, $C_{\nu_i,V/A}=0$.

Another potential issue is to ensure the precise $(g-2)_{e,\mu}$ measurements are sufficiently unaffected. One must have electron couplings, but these measurements are more sensitive to the axial couplings, so we enforce no axial couplings to the electrons or muons, $C_{e/\mu,A}=0$.

Finally, we do not enforce canonical gauge-invariant Yukawa couplings in the first two generations, but motivated by more natural Yukawa couplings in the third generation, we require that the these charges are gauge invariant, i.e. for the top quark coupling like $(\bar{Q} Y_u \tilde{H} u_R)$, this coupling must have vanishing charge, $-z_{Q_3} - z_H + z_{u_{R_3}} = 0$. We make no further comment on the generation of mass in the first two generations, but appeal to higher scale physics, such as radiative mass generation, or horizontal symmetries. However, our final constraint is that the first two generations be family universal, as we expect whatever mechanism generates the masses should apply in a similar fashion for both the first two generations.

These constraints together entirely fix the 15 fermion charges and SM Higgs charge up to a scale factor, for which we fix the SM Higgs charge to unity. This is detailed in Tab. \ref{tab:charges}.

\begin{table}
\hspace{-0cm}
\begin{minipage}[b]{.6\textwidth}
   \centering
\includegraphics[width=1.0\linewidth]{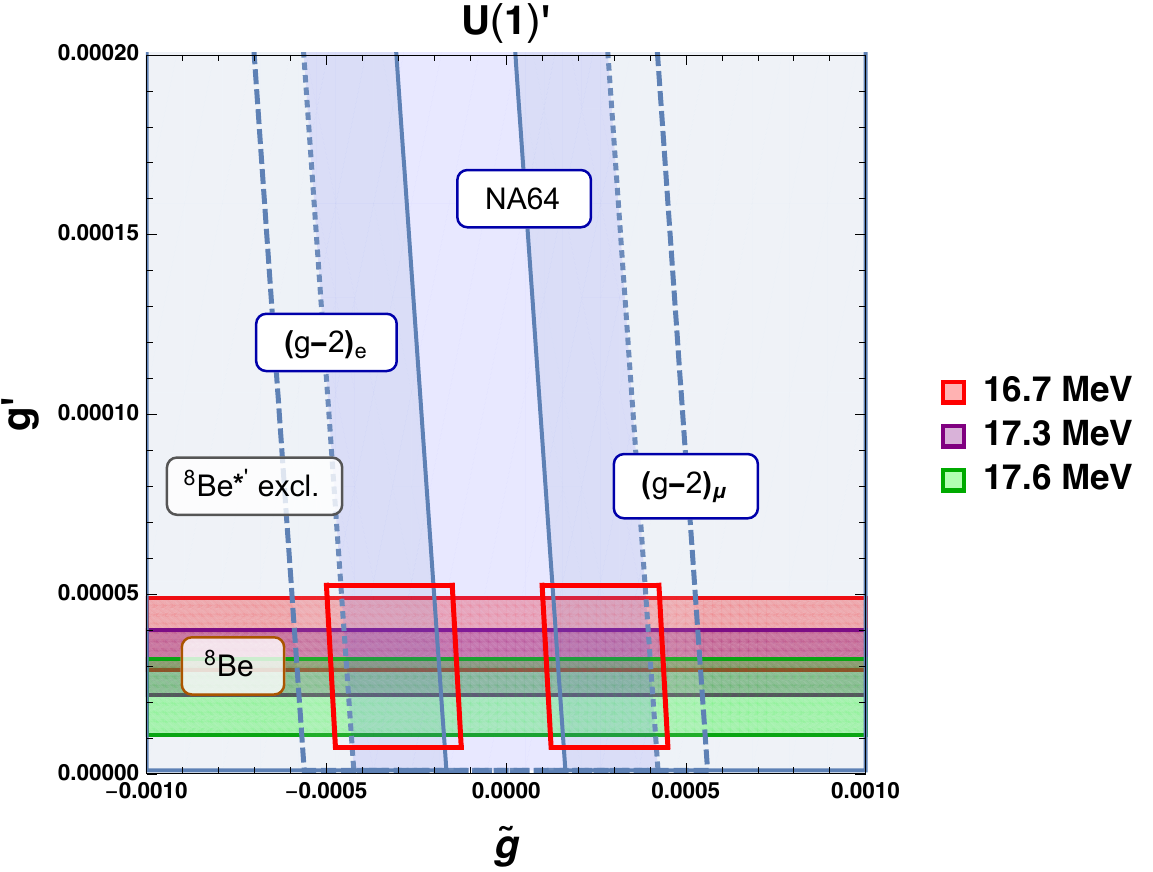}
\captionof{figure}{Allowed parameter space mapped on the $(g',\tilde{g})$ plane 
explaining the anomalous $ \textrm{Be}^*$ decay for $Z'$ solutions with mass 16.7 (red), 17.3 (purple) and 17.6 (green) MeV. The white regions are excluded by the non-observation of the same anomaly in the $ \textrm{Be}^{*'}$ transition. Also shown are the constraints from $(g-2)_\mu$, to be within the two dashed lines; $(g-2)_e$, to be inside the two dotted lines (shaded in blue) and the electron beam dump experiment, NA64, to be in the shaded blue region outside the two solid lines. The surviving parameter space lies inside the two red parallelograms at small positive and negative $\tilde{g}$ (though not at $\tilde{g}=0$), inside the dark shaded blue region which overlaps with the coloured bands, representing the different possibly Atomki anomaly solutions.}
\label{fig:1HDM_ZPHI_0.5_ZQ3_-1Region}
\end{minipage}\qquad
\begin{minipage}[b]{.35\textwidth}
\centering
\resizebox{\columnwidth}{!}{
		\begin{tabular}{|c|c|c|c|c|}
			\hline
			& \multirow{2}{*}{$SU(3)$} & \multirow{2}{*}{$SU(2)$} & \multirow{2}{*}{$U(1)_Y$} & \multirow{2}{*}{$U(1)'$} \\
			&&&&\\ \hline \vspace{-1em}
			&&&&\\
			$Q_{1}$	&  3		&	2	& 1/6	&	$1/3$  \\
			$Q_{2}$	&  3		&	2	& 1/6	&	$1/3$  \\
			$Q_{{3}}$	&  3		&	2	& 1/6	& $1/3$ \\
			$u_{R_{1}}$	&  3		&	1	& 2/3	&	$-2/3$ \\
			$u_{R_{2}}$	&  3		&	1	& 2/3	&	$-2/3$ \\
			$u_{R_{3}}$	&  3		&	1	& 2/3	&	$4/3$  \\
			
			$d_{R_{1}}$	&  3		&	1	& -1/3	&	$4/3$ \\
			$d_{R_{2}}$	&  3		&	1	& -1/3	&	$4/3$\\
			$d_{R_{3}}$	&  3		&	1	& -1/3	&	$-2/3$ \\
			$L_{1}$		&  1		&	2	& -1/2	&	$-1$ \\
			$L_{2}$		&  1		&	2	& -1/2	&	$-1$  \\
			$L_{3}$		&  1		&	2	& -1/2	&	$-1$ \\
			$e_{R_{1}}$	&  1		&	1	& -1		&	$0$ \\
			$e_{R_{2}}$	&  1		&	1	& -1		&	$0$ \\
			$e_{R_{3}}$	&  1		&	1	& -1		&	$-2$ \\
			$H$	&  1		&	2	& 	1/2	&	$1$  \\
			\hline
		\end{tabular}
		}
		\caption{Charge assignment of the SM particles under the family-dependent (non-universal) $U(1)'$. This numerical charge assignment  satisfies the discussed anomaly cancellation conditions, enforces a gauge invariant Yukawa sector of the third generation and family universality in the first two fermion generations as well as  no coupling of the $Z'$ to the all neutrino generations.}
		\label{tab:charges}
\end{minipage}
\end{table}

\section{Results}
There are many experiments which will constrain some areas of parameter space, but for the sake of brevity we discuss only the most constraining ones in our particular parameter space which may satisfy the Atomki anomaly.

As we must have vector couplings to the electrons, and require first and second generation universal couplings, hence also to the muons, there is a contribution to their anomalous magnetic moments, $(g-2)_{e,\mu}$. 

The other experiment which then bounds our entire parameter space is the electron beam dump NA64 experiment. This experiment looks for dark photons which bremsstrahlung from electrons scattered from target nuclei struck by an incoming electron beam dump.

We present our allowed parameter space in the gauge coupling, $g'$, and gauge-kinetic mixing, $\tilde{g}$, plane. We colour the three different mass bands allowed by the three different masses detailed as solutions by the Atomki collaboration. The $(g-2)_\mu$ constrains allowed parameter space to lie inside the dashed blue lines, and similarly for $(g-2)_e$ with a dotted line. The NA64 experiment requires one to lie outside the solid blue lines. Regions of parameter space which can explain the anomaly as well as evade all experimental constraints lie inside the two red parallelograms, in the dark blue shaded region, overlapping the other coloured bands, at small positive and negative (but non-zero) $\tilde{g}$, and at small $g'$.

\section{Conclusion}
In this proceedings, we have presented a viable model to explain the current Atomki anomaly, through a $U(1)'$ extension of the SM, with a new $Z'$ with mass around 17 MeV, and gauge coupling $g' \sim 10^{-5}$ and gauge-kinetic mixing $\tilde{g} \sim 10^{-4}$. Driven by a simple set of requirements, we confine the free SM field charges under the $U(1)'$ to be entirely fixed, and find viable parameter space left which can avoid all constraints.

\section*{Acknowledgements}
The work of LDR and SM is supported in part by the NExT Institute. SM also acknowledges partial financial contributions from the STFC Consolidated Grant ST/L000296/1. Furthermore, the work of LDR has been supported by the STFC/COFUND Rutherford International Fellowship Programme (RIFP). SJDK and SK have received support under the H2020-MSCA grant agreements InvisiblesPlus (RISE) No. 690575 and Elusives (ITN) No. 674896. In addition SK was partially supported by the STDF project 13858. All authors acknowledge support under the H2020-MSCA grant  agreement NonMinimalHiggs (RISE)
 No. 64572.

\end{document}